\documentclass[11pt]{article}
\usepackage{amsmath}
\usepackage{amssymb,amsfonts}
\usepackage{layout}

\def\ni{\noindent}
\def\ga{\gamma}
\def\Ga{\Gamma}
\def\bm{\bar{m}}

\def\nn{\nonumber}

\def\a{\alpha}
\def\at{\tilde{\alpha}}
\def\u{\Upsilon}

\def\h3{$\textrm{H}_3^+$}
\def\bm{b^{-1}}

\def\d{{\partial}}

\def\lbldef#1#2{\expandafter\gdef\csname #1\endcsname {#2}}

\def\href#1#2{#2}

\newcommand{\beq}{\begin{equation}}
\newcommand{\eeq}{\end{equation}}
\newcommand{\ber}{\begin{eqnarray}}
\newcommand{\eer}{\end{eqnarray}}

\def\be{\begin{eqnarray}}
\def\ee{\end{eqnarray}}

\newcommand{\cfa}[4]{\!{\textstyle \left[ {{#1}\atop {#2}} {{#3}\atop
    {#4}} \right] }}

\begin{document}

\title{\flushright \small YITP-SB-05-46 \\
TIFR/TH/05-48 \\
\vskip 2cm
\center
\LARGE
\textbf{
Liouville theory
without an action}
\vskip 1cm
}

\author{\normalsize Ari Pakman\footnote{ari.pakman@stonybrook.edu, apakman@theory.tifr.res.in}}
\date{}
\maketitle
\normalsize
\begin{center}
C.N.Yang Institute for Theoretical Physics, \\
Stony Brook University, \\
Stony Brook, NY 11794-3840, USA

\vskip 0.5cm
and
\vskip 0.5cm
Tata Institute of Fundamental Research, \\
Homi Bhabha Rd., Mumbai 400 005, India
\end{center}


\vskip 2cm

\begin{abstract}
\ni
We show that the crossing symmetry of
the four-point function in the Liouville
conformal field theory on the sphere
contains more information than what
was hitherto considered.
Under certain assumptions,
it provides  the special
structure constants that were previously computed
perturbatively and
allows to solve the theory without
using the Liouville interaction.
\end{abstract}
\thispagestyle{empty}
\newpage


\setcounter{page}{1}
\setcounter{section}{0}
\ni

\section{Introduction}
The constraints imposed by
symmetries on a physical
theory achieve their most powerful realization when they
completely
determine  its dynamics.
A typical example of this occurs
in rational conformal field theories (CFT) in two dimensions
\cite{Belavin:1984vu,Knizhnik:1984nr}, where
the dynamics
of the local fields can be  solved without using
any action, even though in many cases
(as in WZW models for compact groups or unitary minimal models)
the theories have well-defined Lagrangian realizations.

Dynamical constraints on a two-dimensional CFT
follow from the existence of degenerate representations
of the chiral algebra and the
crossing symmetry, i.e.,
associativity of the operator product expansion \cite{Belavin:1984vu}.
In the last years significant  progress was  made
in implementing these
constraints
in interacting non-rational CFTs, i.e., theories with a continuous set of primaries
(for reviews see \cite{Seiberg:1990eb,Teschner:2001rv, Nakayama:2004vk, Schomerus:2005aq}).
In this note we will focus on Liouville theory,
which is the simplest and best understood example
of this class of models,
and is  important  in the study of
non-critical strings~\cite{Polyakov:1981rd}.
Extending previous results, we will see that  the conformal bootstrap program
can be carried another step forward, allowing to compute
the three-point function without using the Liouville interaction at all.

The Liouville theory is defined by the action
\be
S_L[\phi]= \frac{1}{4\pi}\int d^2z \left[\d \phi \bar{\d}\phi + RQ\phi \right]
+  \mu \int d^2z e^{2b\phi} \,.
\ee
It is a CFT with central charge
\be
c= 1 + 6Q^2\,,
\ee
where
\be
Q= b + \frac{1}{b} \,.
\ee
For $b \in \mathbb{R}$, we have $c \geq 25$.
The observables $V_{\a}= e^{2\a \phi}$
have conformal dimension
\be
\Delta_{\a}= \a(Q-\a)\,,
\label{dimension}
\ee
and normalizable states correspond to $\a = \frac{Q}2 + \textrm{i}  \mathbb{R}^+$
\cite{Curtright:1982gt}.
Since these
do not belong to
degenerate representations of the chiral algebra,
not all the methods of~\cite{Belavin:1984vu} can be used.
Early attempts to compute correlators resorted thus to
more conventional
path integral or operator quantizations methods,
based on the action.
In the path integral approach, for example,
correlation  functions are given by\footnote{
We omit the explicit anti-holomorphic dependence of the fields unless needed.}
\be
\langle e^{2\a_1 \phi(z_1)}\ldots e^{2\a_n \phi(z_n)}\rangle &=&
\int {\cal D}\phi \prod_{r=1}^n e^{2 \a_r \phi(z_r)} e^{-S_L[\phi]} \,.
\label{liouville-path}
\ee
It is useful to
split the Liouville
field into a zero mode and fluctuations around it as $\phi=\phi_0 + \tilde{\phi}$.
Performing the formal integral over $\phi_0$ in (\ref{liouville-path})
yields \cite{Goulian:1990qr}
\be
\langle e^{2\a_1 \phi(z_1)}\ldots e^{2\a_n \phi(z_n)}\rangle =
\frac{\Gamma(-s)}{2b}\langle \prod_{r=1}^n e^{2\a_r \tilde{\phi}(z_r)}
\left(\mu \int d^2z e^{2b \tilde{\phi}(z)} \right)^s \rangle_{free}
\label{gl}
\ee
where
\be
s=\frac{Q-\sum_{r=1}^n \a_r}{b} \,.
\label{ese}
\ee
The resulting expression makes sense when $s$ is a non-negative integer.
In this case the factor $\Ga(-s)$ gives a pole associated to the
non-compactness of the target space, and
a free field computation gives the residue.
In particular, when $\sum_{r=1}^{n}\a_r=Q$, we have $s=0$ and no insertions of the
Liouville interaction are required.
 For two vertices, this suggests
adopting the state $\langle Q-\a|$ as the BPZ conjugate of $|\a \rangle$,
since $V_{\a}$ and $V_{Q-\a}$ have the same conformal dimension $\Delta_{\a}$ in (\ref{dimension}), and
\be
\langle Q-\a|\a \rangle = \lim_{z \rightarrow \infty }
|z|^{2\Delta_{\a}} \langle e^{2 (Q-\a) \phi(z,\bar{z})} e^{2\a \phi(0,0)}\rangle = 1 \,,
\label{BPZ}
\ee
up to a divergent factor associated to the non-compactness of the zero mode.

These results led the authors of~\cite{Dorn:1992at, Dorn:1994xn, Zamolodchikov:1995aa}
to propose the following expression for the three point function
\be
\langle V_{\a_1}(z_1) V_{\a_2}(z_2) V_{\a_3}(z_3) \rangle
= \frac{C(\a_1,\a_2,\a_3)}
{|z_{12}|^{2\Delta_1 + 2\Delta_2 - 2\Delta_3}
|z_{23}|^{2\Delta_2 + 2\Delta_3 - 2\Delta_1}
|z_{31}|^{2\Delta_3 + 2\Delta_1 - 2\Delta_2}
}\,,
\ee
with
\be
C(\a_1,\a_2,\a_3) = \left[\pi \mu \ga(b^2)b^{2-2b^2} \right]^{\frac{(Q-2\at)}{b}}
\frac{\u'(0)}{\u(2 \at - Q)} \prod_{r=1}^3 \frac{\u(2 \a_r)}{\u(2\at_r)} \,.
\label{DOZZ}
\ee
We have defined
\be
\at & =& \frac12 (\a_1 + \a_2 + \a_3) \qquad \qquad \at_r = \at - \a_r\,,
\label{tilde}
\\
\ga(x)& =& \frac{\Ga(x)}{\Ga(1-x)}\,,
\ee
and the definition and properties of
the function $\u(x)$ are recalled in Appendix A.

Eq.(\ref{DOZZ}), so-called DOZZ formula, has poles at the expected values of $s$
with the correct residues, and has passed several consistency
tests\footnote{In fact, it has more poles than expected. This is
due to the $b \leftrightarrow \bm$ duality present in the full quantum theory.}.
But the full power of
the conformal symmetry remained hidden
from this important result
until
it was shown
in
\cite{Teschner:1995yf}
how the DOZZ formula
fits into the framework of the
conformal bootstrap of \cite{Belavin:1984vu}.
This approach, dubbed the Teschner trick, led to the computation
of other bulk and boundary quantities in Liouville theory
and in other non-rational CFTs, such as $N=1,2$ Liouville
and the \h3\ WZW model (see \cite {Schomerus:2005aq} for full references).

The method of \cite{Teschner:1995yf}, which we review below, still used the Liouville Lagrangian,
but reduced its role to a minimum.
To obtain the three-point function, for example,
one only needs to compute a special structure constant $C_-(\a)$
appearing in the fusion of the degenerate Virasoro primary
$V_{-b/2}$ with a generic primary~$V_{\a}$:
\be
V_{-b/2}V_{\a} =   C_+(\a) \left[ V_{\a - b/2}\right]
+  C_-(\a) \left[V_{\a +  b/2}\right]\,.
\label{lfusion}
\ee
As follows from (\ref{ese}) and (\ref{BPZ}),
$C_-(\a)$ can be obtained perturbatively
using just one insertion of the Liouville interaction,
\be
C_-(\a)& =&  \langle V_{Q-\a - \frac{b}{2}}(\infty) V_{-\frac{b}{2}}(1) V_{\a}(0)\rangle  \nn \\
&=& -\mu\int d^2x
\langle V_{Q-\a - \frac{b}{2}}(\infty)e^{2b\phi}(x) V_{-\frac{b}{2}}(1) V_{\a}(0)\rangle_{free}
\nn \\
&=& -\mu \int d^2x |x|^{-4\a b}|x-1|^{2b^2} \nn \\
&=& -\pi \mu \frac{\gamma(-1+2\a b -b^2)}{\gamma(2\a b)\gamma(-b^2)} \,,
\label{cminus}
\ee
where in the last line we have used the formula (\ref{usefulintegral}).
The second special structure constant is
\be
C_+(\a)= \langle V_{Q-\a +b/2}(\infty) V_{\a}(1) V_{-b/2}(0)\rangle = 1 \,,
\label{c-one}
\ee
since, according to (\ref{ese}), we need here
no insertions of the Liouville
interaction.\footnote{There is a slight
abuse of notation here.
Following (\ref{gl})-(\ref{ese}),
the r.h.s. of (\ref{cminus}) and (\ref{c-one})
are expected to have a pole, and they
should be read as defining the corresponding residue.
The same holds for similar expressions below.
We will also evaluate
quotients of
$C(\a_1,\a_2,\a_3)$ at divergent values.
In these cases the poles will cancel and we get the quotient of the residues.
\label{footn}
}

\vskip 0.5cm
The main result of this note is that, assuming that
(\ref{BPZ}) and (\ref{c-one}) hold,   we obtain
the special  structure constant $C_-(\a)$
from the conformal
bootstrap, thus allowing to completely solve the theory
without ever using the Liouville interaction.

\section{Exploiting the bootstrap}
The bootstrap program for two-dimensional CFTs
consists in determining the three-point functions from the constraints
imposed by the crossing symmetry of the four-point functions \cite{Belavin:1984vu}.
The original approach to this program, successfully applied
in \cite{Belavin:1984vu,Dotsenko:1984ad} for the minimal models, required the construction
of monodromy invariant combinations of conformal blocks for
Virasoro degenerate fields of arbitrarily high order.

The trick of Teschner \cite{Teschner:1995yf} is an alternative implementation
of the bootstrap idea, appropriate to compute correlators
of non-degenenerate primaries.
Instead of yielding the three-point functions themselves,
the method gives difference equations for them.
The latter must then be  solved to obtain the final answer.
Compared to the original bootstrap  the method is very efficient,
 because for {\it any} three-point function
one only needs the conformal blocks of the first Virasoro degenerate fields,
which are given by  standard hypergeometric functions.

Let us start by recapitulating the argument in \cite{Teschner:1995yf}.
The global conformal symmetry fixes the form of a  four-point function to be
\be
\label{4pf}
\langle V_{\a_4}(z_4) V_{\a_3}(z_3) V_{\a_2}(z_2) V_{\a_1}(z_1) \rangle &=&
G_{\a_4 \a_3 \a_2 \a_1} (\eta,\bar{\eta})
\\
&& \!\!\!\!\!\!\!\!\!\!\!\!\!\!\!\!\!\!\!\!\!\!\!\!\!\!\!\!\!\!\!\!\!\!\!\!\!\!\!\!\!\!\!\!\!\!\!\!\!\!\!\!\!\!\!\!\!\!\!\!\!\!\!\!\!\!\!\!\!\!\!\!\!\!\!\!\!\!\!\!\!
\times \,\, |z_{24}|^{-4\Delta_2} |z_{14}|^{2(\Delta_2+\Delta_3-\Delta_1-\Delta_4)} |z_{34}|^{2(\Delta_1+\Delta_2-\Delta_3-\Delta_4)} |z_{13}|^{2(\Delta_4-\Delta_1-\Delta_2-\Delta_3)}
 \nn
\ee
where
\be
\eta=\frac{z_{12}z_{34}}{z_{13}z_{24}} \qquad
\bar{\eta}=\frac{\bar{z}_{12} \bar{z}_{34}}{\bar{z}_{13}\bar{z}_{24}}
\,.
\ee
Among the non-normalizable states, the pair
\be
\a_{2,1}= -\frac{b}{2}
\qquad
\a_{1,2} = -\frac{1}{2b} \,,
\ee
corresponds to degenerate states with a singular descendant
at level $2$.
To obtain constraints on the  three-point function $C(\a_1,\a_2,\a_3)$,
we consider a four-point function
with $\a_2= \a_{2,1}= -b/2$.
We are assuming that the values of $\a_i$ in the correlators
can be analytically
continued from the normalizable values to non-normalizable ones.
The decoupling of the singular vector
\be
\left[ L_{-2} + \frac{1}{b^2}L_{-1}^2 \right] V_{-b/2}=0 \,,
\label{singeq}
\ee
implies then that $ G_{\a_4 \a_3 \a_2 \a_1}(\eta,\bar{\eta})$ satisfies
\be
\left[ -\frac{1}{b^2}\frac{d^2}{d \eta^2} + \left( \frac{1}{\eta-1} + \frac{1}{\eta} \right) \frac{d}{d \eta}
-\frac{\Delta_3}{(\eta-1)^2} - \frac{\Delta_1}{\eta^2} + \frac{\Delta_1 + \Delta_2 + \Delta_3 -\Delta_4 }{\eta(\eta-1)}
  \right] G_{\a_4 \a_3 \a_2 \a_1} (\eta,\bar{\eta}) \,,
\label{degenerateeq}
\ee
and a similar anti-holomorphic equation.
The solutions to this equation
near $\eta \sim 0,1,\infty$  go
\linebreak
like $\eta^{\Delta_{\a_1 \pm b/2}-\Delta_1 - \Delta_2}$,
$(1-\eta)^{\Delta_{\a_3 \pm b/2}-\Delta_3 - \Delta_2}$ and
$(1/\eta)^{\Delta_{\a_4 \pm b/2}-\Delta_4 - \Delta_2}$, respectively.
From this we see that the fusion of $V_{-b/2}$ with any primary $V_{\a}$ gives
  $V_{\a \pm b/2}$, as in
(\ref{lfusion})\footnote{Note that eq.(\ref{lfusion}) assumes that
there is only one field for every conformal dimension
(the vertices $V_{\a}$ and $V_{Q-\a}$ are related by the reflection
coefficient $R(\a)$ as $V_{\a}= R(\a)V_{Q-\a}$, see \cite{Zamolodchikov:1995aa}).
This property would distinguish between Liouville theory and
a direct sum of several Liouville CFTs with the same total central charge.
}.
Also from (\ref{degenerateeq}) follows that the
conformal blocks can be expressed through hypergeometric functions.
Expanding around $\eta \sim 0$,  we get
\be
G_{\a_4 \a_3 \a_2 \a_1} (\eta) &=&
\sum_{s=+,-}
C(\a_4,\a_3,\a_1 - s b/2) C_s(\a_1)
\left| {\cal F}_s  \cfa{\a_3}{\a_4}{\a_2}{\a_1}
(\eta)\right|^2  \,,
\label{ge}
\ee
where
\be
C_{\pm}(\a) \equiv C(Q-\a \pm b/2,-b/2,\a)\,
\label{cpm}
\ee
are the special structure
constants
in~(\ref{lfusion}).
The conformal blocks are
\be
{\cal F}_+  \cfa{\a_3}{\a_4}{-b/2}{\a_1}(\eta) &=&
\eta^{\Delta_{\a_1 -  b/2}-\Delta_1 - \Delta_2}(1-\eta)^{b \a_3}F(A,B;C;\eta) \,,
\label{cb1}
\\
{\cal F}_-  \cfa{\a_3}{\a_4}{-b/2}{\a_1}(\eta) &=& \eta^{\Delta_{\a_1 +  b/2}-\Delta_1 - \Delta_2}(1-\eta)^{b \a_3}F(A-C+1,B-C+1;2-C;\eta) \,,
\label{cb2}
\ee
with
\be
A &=& -1 + b(\a_1 + \a_3 + \a_4 -3b/2)\,, \\
B&=& b(\a_1 + \a_3 - \a_4 -b/2)\,, \\
C &=& 2 \a_1b -b^2 \,.
\ee

\vskip 0.5 cm
\ni
The interchange of $V_{\a_1}(z_1)$ and $V_{\a_4}(z_4)$
in (\ref{4pf}) leads to the crossing symmetry relation
\be
G_{\a_4 \a_3 \a_2 \a_1} (\eta,\bar{\eta}) &=&
|\eta|^{-4\Delta_2} G_{\a_1 \a_3 \a_2 \a_4} (1/\eta,1/\bar{\eta}) \,,
\label{crosssym}
\ee
which will lead to the  crucial relation to exploit below.
Now, the {\it s}-channel conformal blocks (\ref{cb1})-(\ref{cb2})
can be expressed through the {\it u}-channel conformal blocks
${\cal F}_{\pm} \cfa{\a_3}{\a_1}{\a_2}{\a_4} (1/\eta,1/\bar{\eta})$
by means of the standard identity
\be
F(A,B;C;\eta) &=& \frac{\Gamma(C) \Gamma(B-A)}{\Gamma(B) \Gamma(C-A)} (-\eta)^{-A} F(A,1-C+A;1-B+A; 1/\eta)  \nn \\
&+& \frac{\Gamma(C) \Gamma(A-B)}{\Gamma(A) \Gamma(C-B)} (-\eta)^{-B} F(B,1-C+B;1-A+B; 1/\eta)
\,.
\ee
This leads to a fusing relation of the form
\be
{\cal F}_{s} \cfa{\a_3}{\a_4}{\a_2}{\a_1} (\eta,\bar{\eta})
= \sum_{r= \pm} D_{sr}
{\cal F}_{r} \cfa{\a_3}{\a_1}{\a_2}{\a_4} (1/\eta,1/\bar{\eta})
\,.
\ee
Inserting this transformation into (\ref{ge})
and defining
${\cal F}_{\pm} \equiv {\cal F}_{\pm} \cfa{\a_3}{\a_1}{\a_2}{\a_4} (1/\eta,1/\bar{\eta})$,
we see that (\ref{crosssym}) is equivalent to
\be
\label{bootstrap}
 && C(\a_4,\a_3,\a_1 -b/2) \, C_+(\a_1) \,\, \left| D_{++}{\cal F}_+
 + D_{+-}{\cal F}_-
 \right|^2  \nn \\
& +&   C(\a_4,\a_3,\a_1 +b/2) \,C_-(\a_1)
\,\, \left| D_{-+}{\cal F}_+
 + D_{--}{\cal F}_-
 \right|^2 \nn \\
&=& C(\a_1,\a_3,\a_4 -b/2) \, C_+(\a_4) \,\,
\left| {\cal F}_+  \right|^2
 + C(\a_1,\a_3,\a_4 +b/2) \, C_-(\a_4)
\left| {\cal F}_-  \right|^2 \,.
\ee
This is the main equation to exploit.
First note that by setting to zero the cross terms in the l.h.s. we get
\be
\label{liouville-shift}
\frac{C(\a_4,\a_3,\a_1 +b/2)}{C(\a_4,\a_3,\a_1 -b/2)} &=& -
\frac{C_+(\a_1)}{C_-(\a_1)} \frac{D_{++} \bar{D}_{+-}}{D_{-+}\bar{D}_{--}} \,, \\
&=& \frac{C_+(\a_1)}{C_-(\a_1)} \frac{\gamma(2 \a_1b - b^2)}{\gamma(2 -2 \a_1b + b^2)} \times \nn \\
&\times& \frac{\gamma(b(\a_4 + \a_3 -\a_1) -b^2/2)  }{\gamma(-1-3b^2/2 + b(\a_1 + \a_3 + \a_4))}
\frac{\gamma(1 + b(\a_4 - \a_3 -\a_1) + b^2/2)}{\gamma(-b^2/2 + b(\a_1 - \a_3 + \a_4))}
\,.
\nn
\ee
This is a difference equation for $C(\a_4,\a_3,\a_1)$.
It depends on
the ratio $C_+(\a_1)/C_-(\a_1)$,
which can be obtained
from a perturbative computation, as we did in~(\ref{cminus}) and~(\ref{c-one}).

\vskip 0.5 cm
But it turns out that
the crossing symmetry relation (\ref{bootstrap})
contains more information.
A new equation
can be obtained by plunging (\ref{liouville-shift}) back
into~(\ref{bootstrap}) and considering the
coefficients of $|{\cal F}_-|^2$.
This gives
\be
\frac{C_-(\a_4) C(\a_1,\a_3,\a_4+b/2)}
{C_-(\a_1) C(\a_4,\a_3,\a_1+b/2)} &=&
|D_{--}|^2 - \frac {D_{+-}D_{-+}\bar{D}_{--}}{D_{++}}  \,,
\label{bse}
\\
&=&
\frac{\Ga^2(2+b^2 -2b\a_1)\,\, \Ga^2(2b\a_4 -1-b^2)}
{\Ga^2(b(\a_3+\a_4-\a_1 -b/2)) \, \Ga^2(1+b^2/2 + b(\a_4-\a_3-\a_1))} +
\nn \\
&&
\!\!\!\!\!\!\!\!\!\!\!\!\!\!\!\!\!\!\!\!\!\!\!\!
\!\!\!\!\!\!\!\!\!\!\!\!\!\!\!\!\!\!\!\!\!\!\!\!
\!\!\!\!\!\!\!\!\!\!\!\!\!\!\!\!\!\!\!\!\!\!\!\!
\!\!\!\!\!\!\!\!\!\!\!
+ \, \frac{\Ga^2(-1+2b\a_4 -b^2)
\sin(\pi b(\a_1 -\a_3 +\a_4 -b/2)) \sin(\pi b(\a_1 + \a_3 + \a_4 -b/2 -Q))}
{\pi \ga(-1 + 2b\a_1 -b^2) \ga(b(\a_3 + \a_4 -\a_1 - b/2)) \ga(1+ b^2/2+ b(\a_4 -\a_3 -\a_1)) \sin(\pi b(2\a_1 -b))}\,.
\nn
\ee
This equation is   the  main
results of this note.
No new equations  arise from the coefficients
of~$|{\cal F}_+|^2$ in (\ref{bootstrap}).

The quantum Liouville theory is assumed to
be invariant under the interchange $b \leftrightarrow 1/b$.
Thus the degenerate primary $\a_{1,2}= -\bm/2$
leads to a second pair of functional equations,
obtained from~(\ref{liouville-shift})
and~(\ref{bse}) by replacing $b$ with $1/b$.
Notice that the decoupling equation (\ref{singeq}) for the
$V_{-b/2}$ state can be traced back to the classical
Liouville equation of motion,
but for  $V_{-\bm/2}$ the decoupling equation
has no classical limit, and  is an additional assumption of the
quantum theory.
This is natural in our abstract approach, where the classical
limit or the action play no role and both decoupling
equations stand on the same footing.

We will denote the  special structure constants associated to $V_{-\bm/2}$ as
\be
\tilde{C}_{\pm}(\a) \equiv C(Q-\a \pm \bm/2, -\bm/2,\a)\,.
\ee

\subsection*{Consistency check on the BPZ conjugation}
Let us see first how eq.(\ref{bse}) gives a consistency check on
the  BPZ conjugation (\ref{BPZ}),
by fixing the
quantity\footnote{As in the definition of $C_{\pm}(\a)$,
this expression should be understood as the residue of a pole (see footnote \ref{footn}).
}
\be
N(\a)= C(\a,Q-\a,0)= \langle V_{\a} V_{Q-\a}\rangle\,.
\label{ene}
\ee
Consider  (\ref{bse})
at $\a_1=\a,\a_3=0,\a_4=Q-\a-b/2$.
For these values,
the r.h.s. of eq.(\ref{bse})
becomes~$1$, and
using  $C_-(\a)= C_-(Q-\a-b/2)$,
we get
\be
N(\a+b/2) = N(\a)\,.
\ee
This condition along with a similar one with $b \leftrightarrow 1/b$,
implies that, for incommensurate $b$ and $1/b$,
$N(\a)$ is a constant for {\it any} normalization of the vertex operators.
One usually assumes  that the latter are rescaled so that
\be
N(\a)=1 \,.
\label{nuno}
\ee

\subsection*{The special structure constants and the DOZZ formula}



A relation between the special structure constants $C_{\pm}(\a)$
can also be obtained evaluating eq.(\ref{bse}) at
$\a_1=\a, \a_3=Q-\a, \a_4=-b/2$. This
gives\footnote{In \cite{Teschner:1997ft} it was pointed out that
if we evaluate eq.(\ref{liouville-shift}) at $\a_1=\a_4=\a, \, \a_3=-b/2$,
then we get a relation between $C_-(\a)C_+(\a+b/2)$ and
$C_-(\a-b/2)C_+(\a)$, but this is not enough to
fix $C_-(\a)C_+(\a+b/2)$. Here we obtain its  precise value,
up to the free constant $\mu$.
}

\be
C_-(\a) C_+(\a+b/2)= -
\frac{\pi \mu}{\ga(-b^2)}
\frac{\gamma(-1-b^2+2b\a)}{\gamma(2b\a)}\,,
\label{struc-bootstrap}
\ee
where we have used (\ref{nuno}) and we have defined $\mu$ as
\be
\frac{\pi \mu}{\ga(-b^2)} \equiv  \frac{C_-(-b/2)}{2 \cos(\pi b^2)}
\frac{\Gamma^2(-b^2)}{\Gamma^2(-1-2b^2)} \,.
\label{mu-norm}
\ee
The value of $C_-(-b/2)$ is a free parameter, and
it is convenient to express it through $\mu$.

In (\ref{struc-bootstrap}), an arbitrary election for $C_+(\a)$ determines the
structure constant $C_-(\a)$.
This is  the most we can expect from self-consistency
of the theory, since the structure constants will change
under $\a$-dependent
rescalings of the vertex operators.
Assuming that  the vertex operators are normalized such that
$C_+(\a)=1$, then $\mu$ becomes the
cosmological constant
in the Liouville Lagrangian, and we get for $C_-(\a)$
in (\ref{struc-bootstrap})
precisely the perturbative result
(\ref{cminus})\footnote{An interesting point, which we shall not address here, is
 whether any arbitrary election
for $C_+(\a)$ leads to a consistent theory, and when such an arbitrary election
can be brought to $C_+(\a)=1$ by rescaling the vertex operators.}.
Similar relations hold for the $\tilde{C}_{\pm}(\a)$ structure constants,
and we denote the dual cosmological constant by $\tilde{\mu}$.

Plunging the value of $C_-(\a)$ from~(\ref{struc-bootstrap}) into~(\ref{liouville-shift})
we get the difference equation
\be
\frac{C(\a_3,\a_2,\a_1 +b)}{C(\a_3,\a_2,\a_1)}
&=& -\frac{\ga(-b^2)}{\mu \pi }
\frac{\gamma(2 \a_1b)}{\gamma(1 -2 \a_1b-b^2)}
\nn \\
&& \times \,\,\, \frac{\gamma(b(\a_3 + \a_2 -\a_1-b) )  }{\gamma(b(\a_1 + \a_2 + \a_3-Q))}
\frac{\gamma(1 + b(\a_3 - \a_2 -\a_1))}{\gamma(b(\a_1 - \a_2 + \a_3))} \,,
\label{generalshift}
\ee
and similarly
\be
\frac{C(\a_3,\a_2,\a_1 + \bm)}{C(\a_3,\a_2,\a_1)}
&=& -\frac{\ga(-b^{-2})}{\tilde{\mu} \pi }
\frac{\gamma(2 \a_1\bm)}{\gamma(1 -2 \a_1\bm -b^{-2})}
\nn
\\
 && \times \, \,\, \frac{\gamma(\bm (\a_3 + \a_2 -\a_1-1/b) )  }{\gamma(\bm (\a_1 + \a_2 + \a_3-Q))}
\frac{\gamma(1 + \bm (\a_3 - \a_2 -\a_1))}{\gamma(\bm (\a_1 - \a_2 + \a_3))} \,.
\ee
These functional equations were first obtained by Teschner in \cite{Teschner:1995yf}.
Their solution is given by the DOZZ formula (\ref{DOZZ}),
and $\mu$ and its dual $\tilde{\mu}$ are related
as
\be
\tilde{\mu} \pi \ga(b^{-2})
= \left(\mu \pi \ga(b^2) \right)^{1/b^2}\,.
\ee
The solution is unique for incommensurable $b$ and $\bm$.


\section{Conclusions}
In this note  we have reduced the
dependence of the  Liouville conformal field theory
from its Lagrangian, by showing how the theory
can be solved without using the Liouville interaction at all.
The equations (\ref{BPZ}) and (\ref{c-one})
are assumptions which follow naturally from the
perturbative result (\ref{gl}), but do not involve the interaction.
We believe our  result will be useful to formulate
for a purely constructive
approach to Liouville theory, completely independent
from a local Lagrangian.

\section*{Ackowledgements}
We are grateful to Davide Gaiotto, Gast\'on Giribet and Sunil Mukhi
for helpful conversations.
Special thanks to Joerg Teschner for correspondence and
for his comments on the draft.
We thank the organizers of the
Third Simons Workshop in Mathematics and Physics at Stony
Brook University. This work is supported by
the Simons Foundation.

\appendix

\renewcommand{\theequation}{A.\arabic{equation}}
\setcounter{equation}{0}

\section{Useful formulae}
\ber
\ga(x) &\equiv& \frac{\Gamma(x)}{\Gamma(1-x)} \\
\ga(x)&=& \frac{1}{\ga(1-x)} \\
\ga(x+1) &=& -x^2 \ga(x) \\
\Ga(x) \Ga(1-x)& =& \frac{\pi}{\sin(\pi x)}
\eer

\ber
\int_{\mathbb{R}^2} \!\! d^2 x \,  x^{a} \bar{x}^{\bar{a}} (1-x)^{b} (1-\bar{x})^{\bar{b}} &=&
\pi \frac{\Ga(1+a)}{\Ga(-\bar{a})} \frac{ \Ga(1+b)}{\Ga(-\bar{b})}
\frac{ \Ga(-\bar{a}-\bar{b}-1) }{  \Ga(a+b+2) }
\label{usefulintegral}
\eer
The integral is well defined only when $a-\bar{a}, b-\bar{b} \in Z$.

\subsection*{The function $\u(x)$}
The function $\u(x)$ was introduced in \cite{Zamolodchikov:1995aa}
and can be defined by
\be
\log \u(x) = \int_0^{\infty}
\frac{dt}{t} \left[\left( \frac{Q}{2} -x \right)^2 e^{-t}
+ \frac{\sinh^2(\frac{Q}{2}-x)\frac{t}{2}}{\sinh \frac{bt}{2} \sinh \frac{t}{2b}}
\right]\,.
\ee
The integral converges in the strip $0<\textrm{Re}(x)< Q$.
For other $x$ it is defined by the relations
\be
\u(x + b) = b^{1-2bx}\ga(bx) \u(x)
\qquad
\u(x + 1/b) = b^{-1+2x/b}\ga(x/b) \u(x) \,.
\label{shiftu}
\ee
From these last equations it follows that $\u(x)$ has zeros
at
\be
x &=& (m+1)b +  \frac{(n+1)}{b}\,, \\
x &=& - m b - \frac{n}{b}\,,
\ee
for $m,n$ non-negative integers.

\end{document}